\documentstyle[aps,prb,multicol]{revtex}
\input {epsf.sty}
\begin{document}
\draft
\title{The Pseudogap in YBa$_2$Cu$_3$O$_{7-\delta}$ from NMR in High Magnetic Fields}
\author{V. F. Mitrovi{\'c}, H. N. Bachman,  W. P. Halperin}
\address{Department of Physics and Astronomy,\\
   Northwestern University, Evanston, Illinois 60208}
\author{A. P. Reyes, P. Kuhns, W. G. Moulton}
\address{National High Magnetic Field Laboratory
   Tallahassee, Florida 32310}
%\date{Received 15 April 2000}
\date{Version \today}
\maketitle
\begin{abstract}
We report $^{17}$O(2,3) and $^{63}$Cu(2) spin-lattice relaxation
rates  and the $^{17}$O(2,3) spin-spin relaxation
rate   in different magnetic fields in YBa$_2$Cu$_3$O$_7$ near $T_c$.
Together these measurements enable us to test the magnetic field dependence of the 
pseudogap effect on the spin susceptibility
in different regions of the Brillouin  zone 
using the known form factors for different nuclei as filters.   
Thus, we  study  the momentum dispersion of the 
pseudogap  behavior.
We find that near the antiferromagnetic wave vector the pseudogap is
insensitive to magnetic fields up to 15 T.  In the remaining
region, away from the $(\pi,\pi)$ point, the pseudogap shows a magnetic field dependence 
at fields less then  10 T.
The first result is indicative of the opening of a 
spin-pseudogap that suppresses antiferromagnetic
correlations below a temperature $T^*$; whereas, the second result shows the effect of
pairing fluctuations on the spin susceptibility as a precursory effect of
 superconductivity.
\end{abstract}
\pacs{PACS numbers: 74.25.Nf, 74.40.+k, 74.72.Bk}
%\narrowtext
\vspace{-11pt}

\newcommand{\lrule}{ \end{multicols} \noindent
  \rule{0.5\textwidth}{0.1mm}\rule{0.1mm}{3pt}\newline }
\newcommand{\rrule}{ \noindent \parbox{\textwidth}{
  \hfill\rule[-3pt]{0.1mm}{3pt}\rule{0.5\textwidth}{0.1mm}}
  \begin{multicols}{2} }

\begin{multicols}{2}

%\tableofcontents

%%%%%%%%%%%%%%%%%%%%%%%%%%%%%%%%%%%%%%%%%%%%%%%%%%%%%%%%%%%%%%%%%%%
%%%%%%%%%%%%%%%%%%%% INTRODUCTION  %%%%%%%%%%%%%%%%%%%%%%%%%%%%%%%%
%%%%%%%%%%%%%%%%%%%%%%%%%%%%%%%%%%%%%%%%%%%%%%%%%%%%%%%%%%%%%%%%%%%

\section{Introduction}
\label{Intro}

The nature of the onset of superconductivity in high temperature
superconductors (HTS)  is of considerable interest since it reflects a complex 
interplay between magnetism and superconductivity that
is not yet understood.   Experiments show\cite{Timusk99} that 
below a temperature $T^*$ that is higher than $T_c$, a gaplike structure  
appears in the electronic  excitation spectrum. However, 
at present there is no consensus concerning the
relationship between this pseudogap  and superconductivity
\cite{Ding96,Krasnov00,Renner98,mitrovic99,gorny99}.  There are several possibilities 
which can  be crudely divided into two groups: pairing correlations above
 $T_c$, with a relevant energy scale $k_B(T-T_c)$ and high
energy mechanisms, on the scale $k_BT^*$, such as charge or spin gaps. Measurements under strong
magnetic fields  may help to discriminate between  these and to find correlations between them and
superconductivity.  This idea lead to a series of  NMR spin-lattice relaxation rate measurements 
in high magnetic fields \cite{mitrovic99,gorny99,Zheng99,Zheng00}. 
Recent neutron scattering experiments \cite{dai00} have also investigated  magnetic field effects in
the normal state of HTS. NMR may be particularly useful for investigating
the energy scale of the pseudogap if performed 
over a wide range of  magnetic fields. In addition, we find that NMR 
can probe the $q$ (momentum transfer wave vector) dependence of the 
pseudogap in the spin excitation spectrum by taking advantage of known 
$q$-dependent form factors that are different for various
relaxation experiments with the different nuclei, copper and oxygen. 

Recent NMR experiments that investigate the effect of magnetic field on the pseudogap in nearly
optimally doped YBCO  include measurement of the $^{63}$Cu(2) NMR spin lattice relaxation rate,
$T_1^{-1}$,  by Mitrovi{\'c} {\it et al.}\cite{mitrovic99}
and  Gorny {\it et al.}\cite{gorny99}. The results of these two papers on
  $^{63}$Cu(2) $T_1^{-1}$ are contradictory. 
Gorny {\it et al.}\cite{gorny99} reported that the spin-lattice 
relaxation rate in YBCO$_{7-\delta}$  is magnetic field  independent
indicative of the opening of a spin-pseudogap.  These authors point out
that their results are inconsistent with other reports \cite{mitrovic99,borsa92,carretta96}
which concluded that there is a small but significant magnetic field dependence to $(T_1T)^{-1}$
near $T_c$ in optimally doped material with an interpretation in terms of pairing 
fluctuations \cite{mitrovic99,carretta96}. 
In the present work we confirm the results of Gorny {\it et al.}\cite{gorny99}
for copper relaxation and extend this to $^{17}$O(2,3) experiments that give additional 
insight regarding the onset of
superconductivity.

In this work, we report a complete set of NMR relaxation measurements: $^{17}$O(2,3)  
spin-lattice relaxation rate, $^{17}T_1^{-1}$; $^{63}$Cu(2), $^{63}T_1^{-1}$; and
 the $^{17}$O(2,3) spin-spin relaxation  rate, $^{17}T_2^{-1}$, 
as a function of magnetic field near $T_c$, up to
23 T. These measurements reveal a field dependence of the dynamic spin susceptibility, $\chi(q,\omega
\to 0) =  \chi' + i\chi''$, that varies with $q$.  This indicates that multiple processes of
different origin affect $\chi(q,0)$. Based on $^{63}$Cu NMR experiments Gorny {\it et
al.}\cite{gorny99} pointed out that  
$\chi'' (q,0)$ near $q=(\pi, \pi)$  shows no major field dependence 
on the scale of 10 T.  At this position in the 
Brillouin zone $\chi$ is  strongly enhanced by antiferromagnetic 
(AF) spin fluctuations, and so this result suggests that the temperature dependence they observe is
controlled by a much  higher field scale possibly associated with a spin-pseudogap.  
Our experiments reach similar conclusions. In addition we find from $^{17}$O NMR  that  $\chi''
(q,0)$,  away from the $(\pi,\pi)$ point, is magnetic field dependent on the scale of 10 T.
Whereas such behavior might be expected near $q=(0,0)$, it is less clear for momenta in the 
intermediate region between $q = (0,0)$ and $q = (\pi,\pi)$.
This field dependence can be explained in terms of superconducting fluctuations, or
a pairing pseudogap that opens up $\sim 20$ K above $T_c$. 
The existence of the  pairing pseudogap is compatible with 
 a Fermi-liquid like contribution to the susceptibility. 
We describe the experiment in \mbox{Sec. \ref{Experiment}}. 
In \mbox{Sec. \ref{NMRTools}} we discuss how  NMR can be used to probe 
 the $q$-dependent susceptibility. Results and discussion are presented in 
\mbox{Sec. \ref{CUT1Results}-\ref{OT1Results}}.

%%%%%%%%%%%%%%%%%%%%%%%%%%%%%%%%%%%%%%%%%%%%%%%%%%%%%%%%%%%%%%%%%%%
%%%%%%%%%%%%%%%%%%%%   EXPERIMENT   %%%%%%%%%%%%%%%%%%%%%%%%%%%%%%%
%%%%%%%%%%%%%%%%%%%%%%%%%%%%%%%%%%%%%%%%%%%%%%%%%%%%%%%%%%%%%%%%%%%

\section{Experiment}
\label{Experiment}

We have investigated two samples. The first sample, {\it A},  has been used in our previous   
work\cite{mitrovic99,Bachman99} on spin relaxation and Knight shift. It is a near-optimally doped
$\sim 30-40\%$
$^{17}$O-enriched, YBa$_2$Cu$_3$O$_{7-\delta}$, aligned powder sample. 
This sample has a  relatively narrow NQR line width of $\approx$ 290 kHz and was provided 
courtesy of P. C. Hammel at Los Alamos National Laboratory. 
Its NQR frequency is $^{63}\nu_{zz}$ = 31.5 MHz.
The second sample, {\it B}, is a $\sim 60\%$
$^{17}$O-enriched, YBa$_2$Cu$_3$O$_{7-\delta}$, aligned powder sample whose  
NQR line width is $\approx$ 450 kHz and  $^{63}\nu_{zz}$ = 31.2 MHz. 
After $^{17}$O exchange at 550 $^\circ$C, this sample
was annealed at 390 $^\circ$C for a week and consequently might be slightly overdoped.
The crystal $\hat c$-axis of both samples were
aligned with the direction of the applied magnetic field, the $z$-axis.
In \mbox{Fig. \ref{Lines}} we show   the first high frequency  satellite 
of the $^{65}$Cu spectra for each of the two 
\begin{figure}[h]
%%%%%%%%%%%%%%%%%%%   F I G U R E   %%%%%%%%%%%%%%%%%%%%
\centerline{\epsfxsize0.90\hsize\epsffile{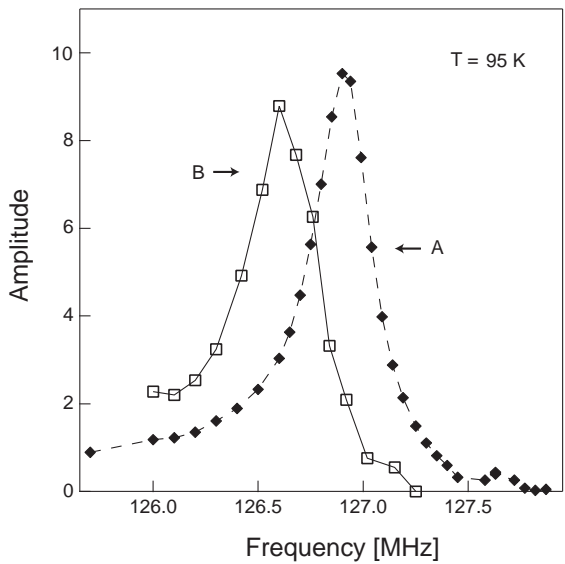}}
%%%%%%%%%%%%%%%%%%%%%%%%%%%%%%%%%%%%%%%%%%%%%%%%%%%%%%%%
\begin{minipage}{0.95\hsize}
\caption[]{\label{Lines}\small 
 The first high frequency satellite, i.e. 
$\left\langle {-{3 \over 2}\leftrightarrow -{1 \over 2}} \right\rangle$ transition, 
spectrum  of  $^{65}$Cu at 8 T and  95 K for
sample {\it A}, solid diamonds, and for sample {\it B}, open squares. The signal is essentially
background free on the high frequency side.
}
\end{minipage}
\end{figure}
\noindent  
samples  at 8 T and 95 K. 
We have checked that the width of these spectra is the same as  
the NQR linewidth. 
Low-field magnetization data, for both samples, show a sharp 
transition at $T_c(0)=92.5\,\mbox{K}$.
Our measurements were made at  temperatures from 70 to 160 K and 
over a  wide range of magnetic fields, from 1.1 T to 22.9 T. 
$^{17}$O(2,3) NMR spin-spin relaxation was
 measured using a Hahn echo sequence: 
$\pi /2$-$\tau $-$\pi $-acquire. Our typical $\pi/2$ pulse 
lengths were 1.5 $\mu $s, except at 2.1 T where pulse lengths 
were 2.5 $\mu $s, giving us a bandwidth $ > 100$ kHz.

The spin-lattice relaxation rate was measured using the 
following sequence: $\pi/2 - \tau_1 - \pi /2-\tau -\pi $-acquire.
$^{17}T_1^{-1}$ was measured on the first high frequency satellite, i.e. 
$\left\langle {-{3 \over 2}\leftrightarrow -{1 \over 2}} \right\rangle$ 
Zeeman transition, of the O(2,3). 
To exclude the possibility of some field dependent background contribution to the rate,
we have compared $T_1^{-1}$ values measured on that satellite to the rate
measured at the
$\left\langle {{3 \over 2}\leftrightarrow {1 \over 2}} \right\rangle $ transition.
$^{63}T_1^{-1}$ was measured using the same sequence as for
measurement of the $^{17}T_1^{-1}$ except that   typical  $\pi/2$ pulse 
lengths were 3 $\mu $s.
All $^{63}T_1^{-1}$ measurements were made on satellites,
$\left\langle {\pm {3 \over 2}\leftrightarrow \pm {1 \over 2}} \right\rangle$.
 At low field, 1.1 and 2.4 T,
the rate was measured on the high frequency satellite of $^{63}$Cu which 
is the highest frequency Cu signal at that field, meaning that the high frequency 
side of this transition is background free, 
following the approach suggested by Gorny {\it et al.}  \cite{gorny99}.
Very good signal-to-noise ratio was obtained even at such low fields
owing to the population difference enhancement by a strong quadrupolar 
interaction.  At 8 T  $T_1$ was measured on the high frequency   
satellite of $^{65}$Cu, the highest frequency Cu signal at that field, whose 
spectrum is shown in \mbox{Fig. \ref{Lines}}.   
The rate of $^{63}$Cu is then inferred 
from $^{65}T_1$ knowing that their ratios scale as the square of 
their gyromagnetic ratios ($\gamma$), namely
${^{63}}T_1 = {^{65}}T_1*({^{63}}\gamma / {^{65}}\gamma)^2 = 0.8713*{^{65}}T_1$. 
At 14.7 T  $T_1$ was measured on the low frequency satellite of  $^{63}$Cu  
 whose low frequency side is background free.
The rates were extracted by fitting to the appropriate recovery profiles, listed in 
\mbox{Table \ref{profT1s}}, assuming a magnetic relaxation mechanism 
(only $\Delta m = \pm 1$ transitions are allowed).

\begin{minipage}{0.93\hsize}
\begin{table}[h]
\begin{tabular}{c|c}
%\hline \\
\\
 Spin &  ${{M_\infty -M(t)} \over {M_\infty }} = $ \\ \\
\hline \\
${3 \over 2}$ & ${1 \over {10}}e^{- t \over T_1} +{1 \over 2}e^{-3t \over T_1}
+{2 \over 5}e^{-6t \over T_1}$ ; \\ \\
${5 \over 2}$  &  ${{2 \over {35}}e^{-t \over T_1}+{3 \over 
{28}}e^{-3t \over T_1}+{1 \over {20}}e^{-6t \over T_1}+{{25} \over
{28}}e^{-10t \over T_1}+{{25} \over {28}}e^{-15t \over T_1}}$\\ \\
\end{tabular}
\begin{minipage}{0.95\hsize}
\caption{\label{profT1s} Recovery profiles for 
$\left\langle {\pm {3 \over 2}\leftrightarrow \pm {1 \over 2}} \right\rangle $ transitions. 
%for spin ${3 \over 2}$, top extression, and ${5 \over 2}$, bottom expresion. 
}
\end{minipage}
\end{table}
\end{minipage}

%%%%%%%%%%%%%%%%%%%%%%%%%%%%%%%%%%%%%%%%%%%%%%%%%%%%%%%%%%%%%%%%%%%
%%%%%%%%%%%%%%%%%%%%%%  NMR TOOLS  %%%%%%%%%%%%%%%%%%%%%%%%%%%%%%%%
%%%%%%%%%%%%%%%%%%%%%%%%%%%%%%%%%%%%%%%%%%%%%%%%%%%%%%%%%%%%%%%%%%%

\section{NMR Tools}
\label{NMRTools}

In this section we give a brief overview of how NMR is used to probe 
the $q$-dependent susceptibility.
The spin-lattice relaxation rate is the rate at which the nuclear
magnetization relaxes to its thermal equilibrium value in the external magnetic field. 
It can be conveniently expressed in terms of 
the generalized spin susceptibility $\chi (q,\omega )$, which is the
response function entering most theoretical descriptions and is also the quantity experimentally
detected by the magnetic part of inelastic neutron scattering. The 
spin-lattice relaxation rate is given by,   
%
%%%%%%%% Equation for T1 %%%%%%%%%%%%%%%%%%%%%%%
\begin{eqnarray}
\label{eqT1}
{1 \over {(T_1T)_\alpha }} \propto \mathop {\lim }\limits_{\omega _n\to 0} \mathop
 \sum\limits_{q,\alpha' \ne \alpha }
 {\left[ {\left| {{}^iF_{\alpha' \alpha' }(q)}
\right|^2{{\chi ''_{\alpha \alpha }(q, \omega _n)} \over {\omega _n}}} \right]}
\end{eqnarray}
\noindent
where $i$ identifies the nuclear species; $\alpha$ is the direction 
of $H_0$ (taken to be parallel to
one of the principal axes of the $F_{\alpha '\alpha '}$ and ${\chi ''_{\alpha '\alpha '}}$ tensors); 
$F_{\alpha '\alpha '}(q)$, referred to as
 a form factor, is the Fourier transform of 
the hyperfine coupling between nuclei and electrons;
and ${\chi ''_{\alpha '\alpha'}(q, \omega)}$ is the imaginary part of the dynamic spin
susceptibility for the wave vector $q$ and nuclear Larmor frequency, $\omega _n$, 
with the direction ${\alpha '}$ perpendicular to $\alpha$. 

The $q$-dependence of relevant form factors in this work, and the imaginary part 
of susceptibility dominated by AF-spin 
fluctuations, are  shown in \mbox{Fig. \ref{FFact}} in \mbox{Appendix \ref{FormFact}} .
We see from \mbox{Eq. (\ref{eqT1})} that it is these form factors 
which  enable us  to probe $\chi (q,\omega _n)$ 
in different  regions of the Brillouin zone, through the measurement of $T_1$.
For $^{63}$Cu(2) spin-lattice relaxation, the appropriate form factor has significant weight near
$q=(\pi,\pi)$, the AF wave vector. 
Since the imaginary part of the susceptibility is peaked at this wavevector the 
copper relaxation is dominated by  AF spin
fluctuations. In contrast for planar oxygen, $^{17}O(2,3)$,  
the spin-lattice relaxation in the normal state is mostly insensitive to AF
fluctuations owing to its vanishingly  small form factor at $q=(\pi,\pi)$.

In most solids, the spin-spin relaxation rate arises from the nuclear
dipole-dipole interaction. This gives rise to a temperature independent  rate, $T_2^{-1}$,  
and decay of the form $\exp (-t^2/2(T_{2G})^2)$ where $(T _{2G})^{-2}$ is equal 
to the second-moment of the homogeneous line-shape (excluding the  broadening due
to the finite lifetime of a spin in an eigenstate). 
However,  nuclei can also interact indirectly via conduction electrons \cite{kittel} 
depending on the real part of their
magnetic susceptibility.  This coupling is an energy conserving process so that 
it contributes to $T_2$.
Thus by measuring $T_2$ one can probe the real part of the electronic susceptibility. 
The importance of $T _{2G}$ of Cu in obtaining information about the AF exchange between 
the electronic spins was first pointed out by Pennington {\it et al.}\cite{pennington89}. 
Whereas these indirect processes dominate the Cu spin-spin relaxation, they are 
strongly reduced for oxygen by its vanishing form factor at $q = (\pi,\pi)$. The 
dominant contribution to oxygen $T_2$ is from direct nuclear dipole-dipole coupling 
between copper and oxygen.
However, an important part of $T _{2G}$ of $^{17}$O(2,3) still arises from 
Cu-O indirect coupling and can be  written as,
\begin{eqnarray}
\label{T2indEq}
& \displaystyle
 \left( {{1 \over {^{63-17}T_{2G}}}} \right)^2_{ind}\propto \sum\limits_q 
\left[{{}^{17}F_{\alpha'}(q)\cdot
 {}^{63}F_{\alpha'}(q)\cdot\chi' (q,0)} \right]^2
\end{eqnarray}
\noindent
where $^{17}F_{\alpha'}(q)$ and $^{63}F_{\alpha'}(q)$ are form factors of O and Cu respectively for 
$\alpha' = c$ for the case $ \hat c || \hat z$.
Unlike the case of $(^{63}T _{2G})_{ind}$ which arises from Cu-Cu indirect coupling 
and probes $\chi' (q,0)$ near $(\pi, \pi)$,
($^{63-17}T _{2G})_{ind}$ arises from Cu-O coupling and  
probes $\chi' (q,0)$ in the intermediate region of
the Brillouin zone between  $(\pi,\pi)$ and $(0,0)$. This relaxation experiment is 
complementary to the measurements of spin-lattice
relaxation.  Finally, the Knight-shift probes the  real-part of static 
spin susceptibility at $q=0$, $\chi' (0,0)$ which we have reported
earlier\cite{Bachman99} for sample {\it A} using a wide range of magnetic fields.

To summarize, in order to characterize the dynamic spin susceptibility at different $q$,
 we have measured the 
following quantities: 
\begin{itemize}
\item{$^{63}T_{1}^{-1}  \propto \chi''/\omega$ for $q$ near $(\pi,\pi)$,}
\item{$(^{63-17}T_{2G})_{ind}^{-1} \propto \chi'$ for $q$ between $(0,0)$ and $(\pi,\pi)$,}
\item{$^{17}T_{1}^{-1}  \propto \chi''/\omega$ for  $q$ near $(0,0)$.}
\end{itemize}
Using these tools, we  investigate  the response of $\chi(q,0)$ to a magnetic field 
near $T_c$  to determine which processes affect $\chi$.

%%%%%%%%%%%%%%%%%%%%%%%%%%%%%%%%%%%%%%%%%%%%%%%%%%%%%%%%%%%%%%%%%%%
%%%%%%%%%%%%%%%%%%%% CU T1 RESULTS  %%%%%%%%%%%%%%%%%%%%%%%%%%%%%%%
%%%%%%%%%%%%%%%%%%%%%%%%%%%%%%%%%%%%%%%%%%%%%%%%%%%%%%%%%%%%%%%%%%%

\section{$^{63}T_1$ Results}
\label{CUT1Results}

In \mbox{Fig. \ref{GCuT1} and \ref{B1CuT1}} we show $^{63}T_1$ for samples {\it A} and 
{\it B} respectively. 
For both samples, we observe no discernible
field  dependence in the normal state 
within experimental accuracy of $\pm   2 \%$. This result is consistent with that reported 
 by Gorny {\it et al.}\cite{gorny99} Above \mbox{$\sim$ 100 K}, $(T_1T)^{-1}$ can be fitted 
to a Curie-Weiss like relation, $(T_1T)^{-1} \propto T_x/ (T + T_x)$,   
where we obtain $T_x = 103$ K based on our 8 T  data. 
This relation for $(T_1T)^{-1}$ is to be expected if 
it is dominated by AF spin fluctuations \cite{millis90}. The peak in 
$^{63}(T_1T)^{-1}$ is observed at $ T^{*} \sim$ 100 K. 
Reduction of 
\begin{figure}[h]
%%%%%%%%%%%%%%%%%%%   F I G U R E   %%%%%%%%%%%%%%%%%%%%
\centerline{\epsfxsize0.93\hsize\epsffile{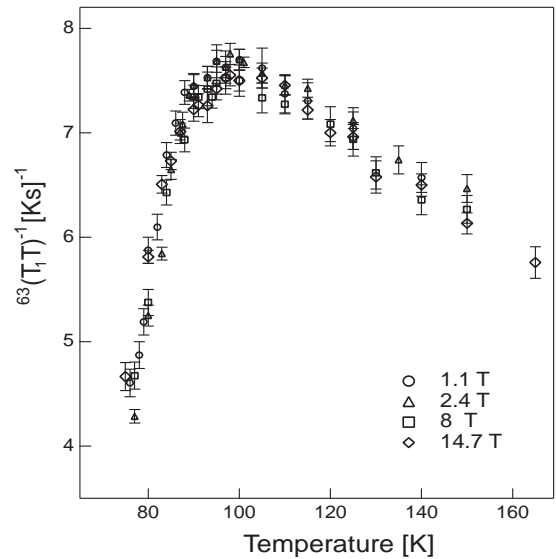}}
%%%%%%%%%%%%%%%%%%%%%%%%%%%%%%%%%%%%%%%%%%%%%%%%%%%%%%%%
\vspace{6pt}
\begin{minipage}{0.95\hsize}
\caption[]{\label{GCuT1}\small
Spin-lattice relaxation rate of $^{63}$Cu(2) in YBCO sample {\it A} as a function of
temperature in the magnetic fields of 1.1, 2.4, 8, and 14.7 T. 
}
\end{minipage}
\end{figure}
\noindent  
$^{63}(T_1T)^{-1}$  below $T^{*}$  has been  associated with  
the loss of low-energy spectral weight\cite{berthier96}, which is caused by the opening of 
a pseudogap. It is interesting to note (for Sample $A$) that in spite of the fact that $T_c$
decreases with field, $^{63}(T_1T)^{-1}$ falls off independently of the magnetic field, 
indicating that down to \mbox{$\sim $ 80 K} the low frequency limit of 
$\chi''(q, \omega)/\omega$ for $q=(\pi,\pi)$ 
is not sensitive to superconductivity and is dominated by a process with a high 
energy scale that becomes gapped and which we refer to as a spin-pseudogap.

\begin{figure}[h]
%%%%%%%%%%%%%%%%%%%   F I G U R E   %%%%%%%%%%%%%%%%%%%%
\centerline{\epsfxsize0.85\hsize\epsffile{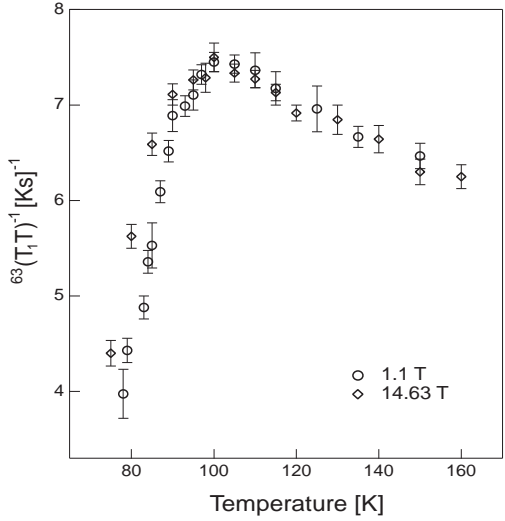}}
%%%%%%%%%%%%%%%%%%%%%%%%%%%%%%%%%%%%%%%%%%%%%%%%%%%%%%%%
\vspace{4pt}
\begin{minipage}{0.95\hsize}
\caption[]{\label{B1CuT1}\small
Spin-lattice relaxation rate of $^{63}$Cu(2) in YBCO sample {\it B} as a function of
temperature in the magnetic fields of 1.1 and 14.63 T. 
}
\end{minipage}
\end{figure}
\noindent  

For sample B the maximum value of $^{63}(T_1T)^{-1}$ 
is shifted to slightly lower values and higher temperatures compared 
to sample $A$. In the superconducting state,   
$^{63}(T_1T)^{-1}$ decreases with decreasing temperature with slight
magnetic field dependence,  not as much as expected 
from $T_c$ reduction by the field \cite{Bachman99}.

In a previous report\cite{mitrovic99} we  inferred $^{63}T_1$ from $^{17}T_2$ 
invoking the theory of Recchia {\it et al.}\cite{recchia96} where we found
significant magnetic field dependence of $^{63}T_1$. 
We now believe that this interpretation  of our $^{17}T_2$ experiment is incorrect as we 
will discuss in the next section.

%%%%%%%%%%%%%%%%%%%%%%%%%%%%%%%%%%%%%%%%%%%%%%%%%%%%%%%%%%%%%%%%%%%
%%%%%%%%%%%%%%%%%%%% CUT1-OT2 RESULTS  %%%%%%%%%%%%%%%%%%%%%%%%%%%% 
%%%%%%%%%%%%%%%%%%%%%%%%%%%%%%%%%%%%%%%%%%%%%%%%%%%%%%%%%%%%%%%%%%%

\section{$^{63}T_1$ from $^{17}T_2$ and Cu-O Indirect Coupling}
\label{CUOIndirect}
Walstedt and Cheong\cite{walstedt95} proposed 
that the main source of spin-echo decay of $^{17}$O is the
copper spin-lattice relaxation. The $z$-component fluctuating fields from copper nuclear 
spin flips are transferred to the oxygen nuclei by Cu-O nuclear dipolar
interactions. 
To account for this process
Recchia {\it et al.}\cite{recchia96} derived an expression for 
$^{89}$Y and  $^{17}$O spin echo height, $M(\tau )$, as a 
function of pulse spacing $\tau $. In order to fully account for the 
$^{17}$O spin-echo decay two additional  mechanisms have to be invoked. 
First is a Redfield contribution\cite{recchia96,slichterB} to the spin-echo decay,
 caused by the finite lifetime  of a spin in an eigenstate as a result of the 
$^{17}$O   spin-lattice  relaxation. This contribution is \mbox{$\sim$ 15 $\%$} at
$T_c$ and can be evaluated using the measured   $^{17}T_1$. The Redfield
contribution is taken to be magnetic field independent and, even if we introduce a weak
magnetic field dependence to this contribution, our fit results do not significantly 
change. The second contribution is an indirect Cu-O nuclear coupling, $k$,
 mediated by the conduction electrons\cite{walstedt95,recchia96}. This effect was
 assumed by Recchia {\it et al.}\cite{recchia96}  and ourselves\cite{mitrovic99}
to be a {\it temperature and field independent} enhancement of the effective Cu-O dipolar
coupling strength.
The following expression from Recchia {\it et al.}\cite{recchia96} gives the
$^{17}$O spin echo height, $M(\tau)$ as a function of pulse  spacing
$\tau$,
\lrule
\begin{eqnarray}
\label{eq1}
& \displaystyle
M=M_0 e^{ \left( -^{17}\! \! \gamma^2 k^2 \sum_{i=1}^\nu \left[ 
\frac{^{63,65}\gamma \hbar }{r_i^3} (1-3\cos^2 \theta_i ) \right]^2 
\times 
\frac{I(I+1)}{3}
\left(T_1^{(i)}\right)^2 \Big[ 2\tau /T_1^{(i)}+
4e^{-\tau/T_1^{(i)}} - e^{-2\tau/T_1^{(i)}}-3 \Big]
- 2\tau/T_{2R} \right)}.
\end{eqnarray}
\rrule

In our earlier work we performed a nonlinear least squares fit of our data to \mbox{Eq. 
(\ref{eq1})}  with $^{63}T_1$  as a fitting parameter choosing $k = 1.57$ to match 
the high temperature results. 
The sum was performed over all Cu neighbors in a 
radius of 12 \AA; $r_i$ is the
Cu-O distance; $\theta_i$ is the angle between the applied field 
and the Cu-O axis; $T_1^{(i)}$ is 
$T_1$ of the $i^{\mbox{th}}$ copper nucleus; $I=3/2$
is the copper nuclear spin;
and $T_{2R}$ is the Redfield contribution to the rate.

We now examine in more detail the relaxation described by  the parameter $k$.
We can extract that part of the spin-spin relaxation due  to Cu-O {\it indirect} coupling  
from our  $^{17}T_2$ data by dividing our measured signal $M$ by that calculated for 
{\it direct} dipolar coupling using \mbox{Eq.
(\ref{eq1})} with $k$ set to 1.  We take into account the relaxation from unmediated 
Cu spin-flips using our direct measurements of
$^{63}T_1^{-1}$.  We then approximately fit the residual decay with a gaussian function of time  
and show the resulting relaxation times in \mbox{Fig. \ref{T2WCDv}} versus temperature for magnetic
fields from 2.1 to 22.8 T. 
There is a well-defined field dependence for $T < 120$ K. 
The qualitative behavior of the rate resembles  the behavior 
observed in the $^{17}$O Knight shift \cite{Bachman99}, a quantity proportional 
to $\chi'(0,0)$, although we do not intend special significance by this comparison. 
Above $T_c$,  $T > 90$ K, there is a
small but clear dependence on field as reported earlier by Mitrovi{\'c} {\it et al.}
\cite{mitrovic99} using a different interpretative framework.  The relatively low 
field scale for this dependence, in contrast to
$^{63}(T_1T)^{-1}$ in
\mbox{Fig. \ref{GCuT1} and
\ref{B1CuT1}}, suggests a connection to superconductivity, most likely from pairing fluctuations.

\begin{figure}[h]
%%%%%%%%%%%%%%%%%%%   F I G U R E   %%%%%%%%%%%%%%%%%%%%
\centerline{\epsfxsize0.90\hsize\epsffile{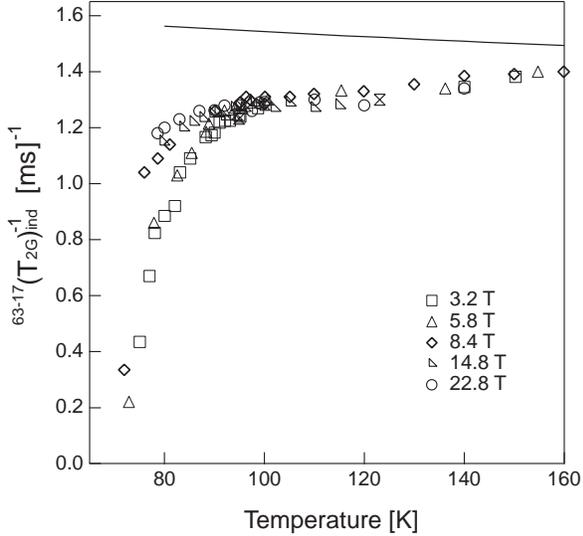}}
%%%%%%%%%%%%%%%%%%%%%%%%%%%%%%%%%%%%%%%%%%%%%%%%%%%%%%%%
\vspace{4pt}
\begin{minipage}{0.95\hsize}
\caption[]{\label{T2WCDv}\small 
Spin-spin relaxation rate of $^{17}$O(2,3) (sample $A$) after dividing out
the part of the relaxation coming from the direct Cu-O dipolar interaction as described in 
the text. The solid
line is the calculated spin-spin relaxation from Cu-O indirect coupling 
using the same susceptibility parameters as we used to calculate $^{17}(T_1T)^{-1}$ and  
$^{63}(T_1T)^{-1}$ discussed later in \mbox{Sec. \ref{OT1Results}}. }
\end{minipage}
\end{figure}  
\noindent 

 Below $T_c$,   the spin-spin relaxation rates  shift to lower temperatures as the field
increases,  consistent with reduction of $T_c$ by the
field,  indicating that this lower temperature behavior is also connected to superconductivity. 
For example, at the applied field of 3.2 T and at the temperature $T$ = 0.9 $T_c(H)$ the rate drops 
by $\sim 20 \% $ from its value at $T_c(H)$. For higher applied fields the decrease is smaller. 
Superconductivity can
affect  $T_2$ data in two ways: through vortex vibrations, whose precise contribution to the rate 
is not known; and, through the suppression of $\chi'$ due to pair-formation. We have shown in 
previous work that vortex vibrations give
rise to a sharp increase in spin-spin relaxation with a lorentzian spectral density that 
onsets at the vortex melting
transition (at least in low fields, $H < 10 T$) \cite{Bachman98}. 
So it seems unlikely that vortex  vibrations are  primarily responsible for the 
field dependence we report in \mbox{Fig. \ref{T2WCDv}} at relatively higher 
temperatures.  To study the effect of pairing on suppression of 
the indirect interaction we have calculated both the
temperature dependence of
$\chi'$ and 
$(^{63-17}T_{2G}^{-1})_{ind}$ in the superconducting state following the RPA-like approach of  
Balut and Scalapino\cite{bulut91}. We take an RPA form for $\chi$,
\begin{eqnarray}
\label{RPAChi}
\chi (q,\omega)={{\chi _0(q,\omega)} \over {1-J_q\chi _0(q,\omega)}},
\end{eqnarray}
\noindent
where $J_q = - J_0(cosq_x + cos q_y)/2$ is an enhancement factor. The value of 
$J_0$ was constrained by matching
$(^{63}T_{2G}^{-1})_{ind}$ experiments from Cu NQR, measurements that are 
unaffected by vortices since they are at zero field.
Below  $T_c$ the measured Cu-Cu indirect coupling contribution to relaxation, 
$(^{63}T_{2G}^{-1})_{ind}$,  decreases by $\sim$ 10 - 20\%  
of its value at $T_c$ \cite{itoh92}. We have reproduced this result 
with our calculation as shown in 
\mbox{Fig. \ref{T2_RPA}}. We find that the  reduction of $(^{63}T_{2G}^{-1})_{ind}$  
is  larger when realistic Fermi surface parameters are taken into account that include  
Fermi surface nesting near the nodes which 
produces  incommensurate peaks in the susceptibility.  We have  calculated
$(^{63-17}T_{2G}^{-1})_{ind}$ using the same parameters for $\chi(q)$ and we find that 
in the superconducting state it decreases by $\sim$ 5 - 10\% of its value at $T_c$, as 
shown in \mbox{Fig. \ref{T2_RPA}}. 
It is conceivable that pairing fluctuations could modify this rate giving raise 
to field dependence above $T_c$. However, this effect is probably too small to account for the 
observed field dependence shown in \mbox{Fig. \ref{T2WCDv}}.

%Thus, we  conclude that only  $\sim$ 5 - 10\%  of decrease of 
%$(^{63-17}T_{2G}^{-1})_{ind}$  with decreasing temperature below $T_c$
%can be ascribed to the suppression of $\chi'$.  

%
\begin{figure}[h]
%%%%%%%%%%%%%%%%%%%   F I G U R E   %%%%%%%%%%%%%%%%%%%%
\centerline{\epsfxsize0.95\hsize\epsffile{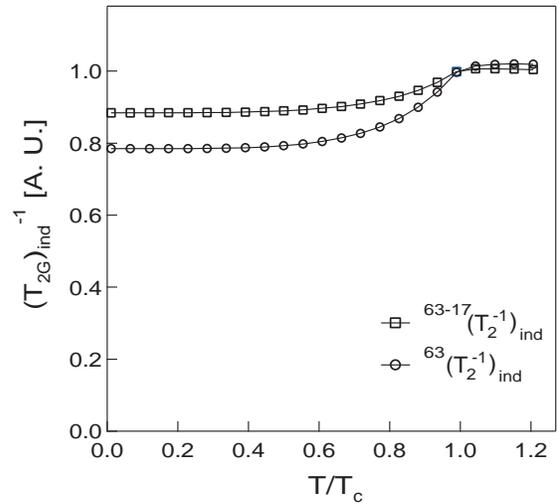}}
%%%%%%%%%%%%%%%%%%%%%%%%%%%%%%%%%%%%%%%%%%%%%%%%%%%%%%%%
%\vspace{4pt}
\begin{minipage}{0.95\hsize}
\caption[]{\label{T2_RPA}\small 
Calculated $(^{63}T_{2G}^{-1})_{ind}$, arising from  Cu-Cu indirect coupling, 
(open circles) and  $(^{63-17}T_{2G}^{-1})_{ind}$, arising from  
Cu-O indirect coupling, 
(open squares) as s function of the reduced temperature, $T/T_c$, for a 
$d$-wave superconductor and $J_0 = 10$.}
\end{minipage}
\end{figure}  
\noindent 

Well above $T_c$ we see in \mbox{Fig. \ref{T2WCDv}} that 
$(^{63-17}T_{2G}^{-1})_{ind}$ decreases as a 
function of decreasing temperature. This appears to be in contrast with our 
calculation of the rate using a phenomenological form for
the susceptibility as discussed in the next section. Nonetheless the magnitude 
of the observed effect is similar to the calculation. One
might also argue that it is possible that the relaxation described by 
$k$,  does not arise only from Cu-O indirect coupling but comes  rather 
from an additional relaxation mechanism which is highly sensitive to 
superconductivity and  associated only with oxygen. 
A possible candidate for this extra  relaxation component is 
the low frequency mostly oxygen charge fluctuations discussed by 
Suter {\it et al.}\cite{Suter00}.  They showed  in YBa$_2$Cu$_4$O$_8$ that  
there is a significant contribution from
quadrupolar fluctuations, i.e. low-frequency charge  fluctuations, 
to $^{17}T_1$ in addition to  the dominant contribution from magnetic
fluctuations. In addition they found evidence that these fluctuations are 
associated with superconductivity.  It might be that these
fluctuations also have spin character and are sensitive to the magnetic field 
providing a possible channel for spin-spin  relaxation. 

Regardless of the precise origin of the  relaxation mechanism described by $k$,  we 
see that it depends on temperature and magnetic field, in contrast with 
previous assumptions\cite{recchia96}.

%%%%%%%%%%%%%%%%%%%%%%%%%%%%%%%%%%%%%%%%%%%%%%%%%%%%%%%%%%%%%%%%%%%
%%%%%%%%%%%%%%%%%%%% O T1 RESULTS  and Discussions %%%%%%%%%%%%%%%%
%%%%%%%%%%%%%%%%%%%%%%%%%%%%%%%%%%%%%%%%%%%%%%%%%%%%%%%%%%%%%%%%%%%
%
\section{$^{17}T_1$ Results}
\label{OT1Results}

As previously pointed out $^{17}(T_1)^{-1}$ probes the imaginary
part of electronic spin susceptibility, $\chi''(q,0)$, close to $q=0$. 
In \mbox{Fig. \ref{GOT1}}
we show the $^{17}$O spin-lattice relaxation rate, of the 
{\it A} sample,  as a function of temperature in different 
magnetic fields.  We find that the rate increases  
with increasing magnetic field, on the scale of 10 T, for $T < 110$ K.  
At 95 K $^{17}(T_1T)^{-1}$  differs by $\sim 7\%$ between  3.2 and 8 T.  
The departure of $^{17}(T_1T)^{-1}$ from the Korringa-like behavior, 
$(T_1T)^{-1} =$ constant, shifts towards  lower temperatures as
the field increases and the rate drops sharply in the superconducting state,
consistent with reduction of $T_c$ by the field \cite{Bachman99}.
Thus, we can conclude that the pseudogap we 
observe here is tied, at least in part, to superconductivity.
A simple shift in $T_c$ is not enough to account for this field dependence
above $T_c$ because the curvature of the data changes with field. 

\begin{figure}[h]
%%%%%%%%%%%%%%%%%%%   F I G U R E   %%%%%%%%%%%%%%%%%%%%
\centerline{\epsfxsize0.95\hsize\epsffile{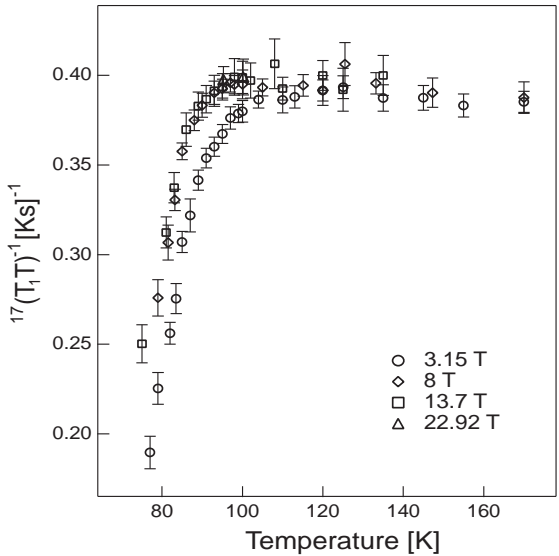}}
%%%%%%%%%%%%%%%%%%%%%%%%%%%%%%%%%%%%%%%%%%%%%%%%%%%%%%%%
\begin{minipage}{0.95\hsize}
\caption[]{\label{GOT1}\small
Spin-lattice relaxation rate of $^{17}$O(2,3) in YBCO (sample $A$) as a 
function of temperature in the magnetic fields of 3.15, 8, 13.7, and 22.92 T.
}
\end{minipage}
\end{figure}
\noindent
Our Knight shift data \cite{Bachman99} indicate a $T_c$ shift 
of $\sim 2$ K from 3.2  to 8 T. 
However,  $^{17}(T_1T)^{-1}$  has a value of 0.367 $\mbox{(Ks)}^{-1}$ 
at 3.2 T at 95 K and at 8 T  the same value at 86 K.
This shift of 9 K exceeds by far the shift of $T_c$  with field.

We can  account for this behavior by 
$d$-wave density-of-states (DOS) pairing fluctuations
following  previously reported analysis \cite{mitrovic99,Bachman99,Eschrig99}.
As the magnetic field increases it suppresses the negative DOS pairing fluctuation  
contribution to the rate causing the  overall rate to increase with increasing 
field \cite{Eschrig99} as observed in $\mbox{Fig. \ref{GOT1}}$. 
Once the fluctuations are completely suppressed by the field \cite{Eschrig99},
we expect the rate to drop sharply at $T_c(H)$. However, we observe that the field 
dependence of the rate saturates around 10 T and that at ``high" field, 
\mbox{ $H \ge $ 10 T}, it has 
well-defined curvature near $T_c$ indicating that DOS pairing fluctuations cannot be the 
only process affecting  $^{17}(T_1T)^{-1}$. In the following we try to model 
the influence of a spin-pseudogap on
$^{63}(T_1T)^{-1}$ and, with the same parameters, estimate the effect on  $^{17}(T_1T)^{-1}$. 

We take the MMP \cite{millis90}  phenomenological expression for the dynamical susceptibility,
altered so as to include the incommensurations in the susceptibility peaks at 
$Q_{i} = (\pi \pm \delta,\pi \pm \delta)$ AF wave vectors  \cite{Zha96},
\begin{eqnarray}
\label{Susc_Eq}
& \displaystyle
\chi (q,\omega )=\chi _{AF}+\chi _{FL}=
& \nonumber \\
& \displaystyle
{1 \over 4}\sum\limits_i {{{\alpha \xi ^2\mu _B^2} \over {1+\xi^2(q-Q_i)^2
-i\omega /\omega _{SF}}}} + {{\chi _0} \over {1-i\pi \omega /\Gamma }}.
& \! \! \! \! \! \! \! \!
\end{eqnarray}
where  $\xi$ is the  spin fluctuation correlation length in units of the lattice 
constant $a$, $\alpha$ 
is a scaling factor, $\omega_{SF}$  
the frequency of spin fluctuations, and   $\xi_0$ and $\Gamma$ are terms added to describe  
 the  Fermi-liquid background for AF fluctuations.

The  imaginary part of $\chi(q,\omega)$ divided by frequency in the limit 
of $\omega \rightarrow 0$ is given by,  
\begin{eqnarray}
\label{ChiIM_MMPEq}
\mathop {\lim }\limits_{\omega \to 0} \mathop
 {\chi''}(q,\omega)/\omega={1 \over 4}\sum\limits_i {{{\alpha \xi (T)^2\mu _B^2/
\omega _{SF}} \over {\left[ {1+\xi (T)^2(q-Q_i)^2}
\right]^2}}}+{{\chi _0\pi } \over \Gamma } .
\end{eqnarray}
The rate   divided by the temperatures, for $H_0||\hat c$ is then evaluated by
summing the product of the form factor and $\chi''$ over all $q$, 
\begin{eqnarray}
\label{T1_MMPEq}
{1 \over {T_1T}} = \mathop {\lim }\limits_{\omega \to 0} \mathop {{k_B} \over {2\mu_B^2\hbar^2}}
\sum\limits_q{F_c(q) {{{\chi''}(q,\omega)} \over {\omega}}}.
\end{eqnarray}
We take Shastry-Mila-Rice\cite{Mila89} form factors given in 
\mbox{Eq. \ref{FF_Eq}}.
In addition,  $\omega_{SF}$ is assumed to be proportional to $\xi(T)^{-2}$ and 
that   $\xi(T) = \xi_0  [{ T_x / (T_x + T})]^{1/2}$. 
Temperature dependence of $\xi(T)$, $\omega_{SF}$, and other parameters were determined so 
that both calculated $^{63}(T_1T)^{-1}$ and $^{17}(T_1T)^{-1}$  coincide with our data.
Assuming that $Q_{AF}=(\pi \pm 0.1, \pi \pm 0.1)$ we find the following values 
for the parameters used to calculate  $(T_1T)^{-1}$s:
$\xi(T) = 3.07 [ 114 \mbox{ K }  /(114 \mbox{ K }    + T)]^{1/2}$, 
\mbox{$\omega_{SF} = 6.09 *\xi(T)^{-2}$ meV}, \mbox{$\alpha = 14.8$ (eV)$^{-1}$},
 and for the Fermi liquid part, \mbox{${{\chi _0}\pi/{\mu _B^2}\hbar \Gamma } 
= 8.885$  eV$^{-2}$}.

We obtain values of $(T_1T)^{-1}$, for both $^{17}$O and  $^{63}$Cu, shown as 
the solid curve (extending to dashed below 120 $K$)
in 
\mbox{Fig. \ref{OCuPSgap}}. We notice that  $^{17}(T_1T)^{-1}$   increases slightly 
with decreasing temperature similar to  $^{63}(T_1T)^{-1}$ due to the increasing correlation 
length for spin fluctuations, indicating that  $^{17}$O is not 
completely shielded  from the AF spin-fluctuations by its form factor. 

\begin{figure}[h]
%%%%%%%%%%%%%%%%%%%   F I G U R E   %%%%%%%%%%%%%%%%%%%%
\centerline{\epsfxsize0.95\hsize\epsffile{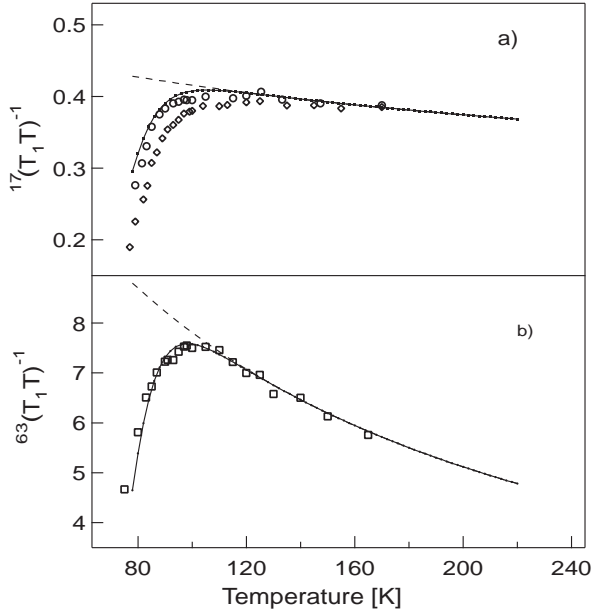}}
%%%%%%%%%%%%%%%%%%%%%%%%%%%%%%%%%%%%%%%%%%%%%%%%%%%%%%%%
\vspace{-6pt}
\begin{minipage}{0.95\hsize}
\caption[]{\label{OCuPSgap}\small
a) Spin-lattice relaxation rate of $^{17}$O(2,3) at 3.15 T (open diamonds) 
and 8 T (open circles);
 and,  b) Spin-lattice relaxation rate of $^{63}$Cu(2) (sample $A$)  at 8 T (open squares) as a 
function of temperature. Solid and dashed lines are 
calculated as explained in the text.
}
\end{minipage}
\end{figure}
\noindent  

We then model the opening of the pseudogap by assuming that it only affects  $\omega_{SF}$. 
We take a phenomenological form for 
$\omega_{SF}^{-1} \propto ([$tanh$((T - T_p)/c_1)]/[  \xi_0^{-2}  \xi(T)^{2} )]$, 
where $c_1 =14.5$ K and $T_p = 70$ K are parameters  
chosen  with the sole purpose to allow a fit to the measured  
 $^{63}(T_1T)^{-1}$, giving the solid curve in \mbox{Fig. \ref{OCuPSgap}b)} below 
$ T \approx 100 K$. Using exactly 
the same pseudogap parameterization, we calculate 
 $^{17}(T_1T)^{-1}$ giving the corresponding solid curve in \mbox{Fig. \ref{OCuPSgap}a)}. We clearly
 see that the suppression of 
$^{63}(T_1T)^{-1}$ due to the opening of the spin-pseudogap, as modeled here, will also cause 
a small suppression of $^{17}(T_1T)^{-1}$ that reproduces the 
observed curvature of the high field $^{17}(T_1T)^{-1}$ data near $T_c$. From our 
simplistic model we have shown phemonenologically
that $^{17}(T_1T)^{-1}$ is affected by the opening of a spin-pseudogap and that 
it is this process that dominates 
the oxygen spin-lattice relaxation rate at fields above 10 T near $T_c$, adding 
to the effects of superconducting pair fluctuations
that give field dependence at low field. 

We have also calculated the rates using  the oxygen form factor suggested by 
Zha {\it et al.} \cite{Zha96}, \mbox{Eq. \ref{FFOnew_Eq}}.    
Results similar  to the ones shown in  \mbox{Fig. \ref{OCuPSgap}} 
were obtained using the following parameters:
$\xi_0 = 2.98$, $T_x = 105$ K, \mbox{$\omega_{SF} = 5.6*\xi(T)^{-2}$ meV}, 
\mbox{${{\chi _0}\pi/{\mu _B^2}\hbar \Gamma }= 13.835$  eV$^{-2}$}
$c_1 =17$ K,  $T_{p} = 69.5$ K. 

We point out that it is possible that pairing fluctuations 
might also affect $\chi''(q,0)$ near $q=(\pi,\pi)$ \cite{Eschrig99}.
However, it is not  observed. 
The observed pairing fluctuation contribution  to $^{17}(T_1)^{-1}$ from
 the Fermi-liquid susceptibility near $T_c$ at small wave vectors, $q$, 
would change $^{63}(T_1)^{-1}$ by less then a percent, 
making it impossible to discern in an experiment. 

\begin{figure}[h]
%%%%%%%%%%%%%%%%%%%   F I G U R E   %%%%%%%%%%%%%%%%%%%%
\centerline{\epsfxsize0.90\hsize\epsffile{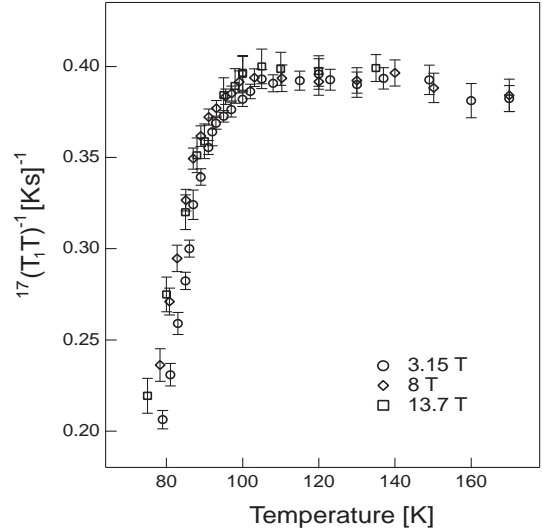}}
%%%%%%%%%%%%%%%%%%%%%%%%%%%%%%%%%%%%%%%%%%%%%%%%%%%%%%%%
\vspace{2pt}
\begin{minipage}{0.95\hsize}
\caption[]{\label{B1OT1}\small
Spin-lattice relaxation rate of $^{17}$O(2,3) in YBCO (sample $B$) as a 
function of temperature in magnetic fields of 3.15, 8, and 13.7 T.
}
\end{minipage}
\end{figure}
\noindent  

In \mbox{Fig. \ref{B1OT1}}
we show the $^{17}$O spin-lattice relaxation rate, of the  {\it B} sample,  
as a function of temperature in different magnetic fields.
At 95 K, $^{17}(T_1T)^{-1}$ differs by $\sim 3\%$ between 3.2 and 8 T, 
which is within the experimental precision) indicating that the field 
dependence of the rate is less then for the {\it A} sample. 
Broader NQR lines for this sample as compared to sample {\it A} 
imply  higher disorder.  We speculate that this would induce more pair-breaking,
ultimately leading to suppression of fluctuations \cite{Eschrig99}.

%%%%%%%%%%%%%%%%%%%%%%%%%%%%%%%%%%%%%%%%%%%%%%%%%%%%%%%%%%%%%%%%%%%
%%%%%%%%%%%%%%%%%%%%%%%      SUMMARY        %%%%%%%%%%%%%%%%%%%%%%%
%%%%%%%%%%%%%%%%%%%%%%%%%%%%%%%%%%%%%%%%%%%%%%%%%%%%%%%%%%%%%%%%%%%

\section{Discussion and Summary}
\label{sumry}

We summarize our relaxation experiments by showing the  relative 
effect of magnetic field  on $\chi(q,0)$ for sample $A$. 
This can be conveniently represented by $R(H)$ defined as
$R(H)$ = $((T_{1,2})_{tot}^{-1} -(T_{1,2})_n^{-1})/(T_{1,2})_n^{-1}$, where 
the normal-state rate, $(T_{1,2})_n^{-1}$, is a fit to the field independent high temperature  
behavior ($T > 120$ K) of the appropriate rate. The results at $T$ = 95 K are 
given in $\mbox{Fig. \ref{Rs}}$. 
The two upper graphs indicate that both 
real and imaginary parts of the spin susceptibility away from 
$q=(\pi,\pi)$ have magnetic field dependence above $T_c$ on the scale of 10 T. 
This field dependence is likely caused by superconducting pair fluctuations that can 
be attributed to the field induced suppression of
the negative  contribution to the rate from the density of states \cite{Eschrig99}. 

\begin{figure}[h]
%%%%%%%%%%%%%%%%%%%   F I G U R E   %%%%%%%%%%%%%%%%%%%%
\centerline{\epsfxsize0.90\hsize\epsffile{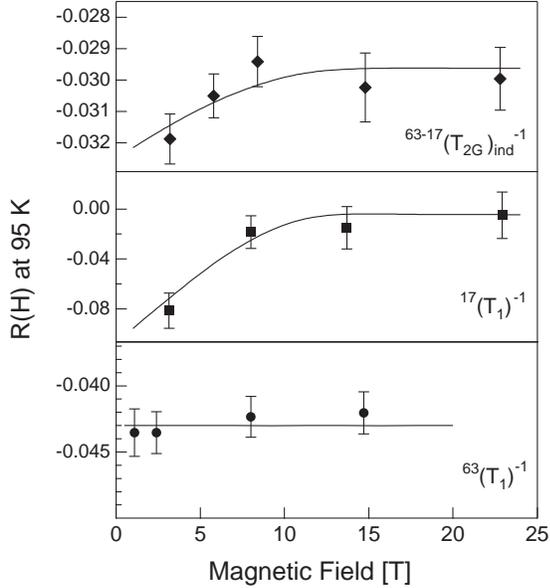}}
%%%%%%%%%%%%%%%%%%%%%%%%%%%%%%%%%%%%%%%%%%%%%%%%%%%%%%%%
\begin{minipage}{0.95\hsize}
\caption[]{\label{Rs}\small
Field dependence of the relaxation 
$R(H)$ = $((T_{1,2})_{tot}^{-1} -(T_{1,2})_n^{-1})/(T_{1,2})_n^{-1}$, 
at 95 K for sample {\it A}.
The normal state rate, $(T_2)_n^{-1}$, of $^{63-17}(T_{2G})^{-1}_{ind}$ 
was defined as $(T_2)_n^{-1} = 1.127 \mbox{(ms)}^{-1} + 
T* 0.00176 \mbox{(Kms) }^{-1}$;  for $^{17}(T_1)^{-1}$ we take 
$(T_1)_n^{-1}  $ = constant =  0.395 \mbox{s$^{-1}$};  and, for $^{63}(T_1)^{-1}$ we take
$(T_1)_n^{-1} =  15.08 \mbox{ s}^{-1}*[104 \mbox{ K }/(104 \mbox{ K } +T)]$.
The solid curves are guides to the eye.
}
\end{minipage}
\end{figure}
\noindent  

 Our measurements show that near $T_c$ the electronic spin susceptibility responds to
a  magnetic field   differently in different parts of the Brillouin zone.
This result  implies  that the spin susceptibility  is affected by different
physical processes. Near $q=(\pi, \pi)$ antiferromagnetic spin 
fluctuations, insensitive to superconducting fluctuations, dominate the spin susceptibility. 
In the   region away from  $q=(\pi, \pi)$  
the susceptibility is influenced by superconducting fluctuations.
 This is consistent with a Fermi-liquid like behavior in which 
the susceptibility is suppressed by superconducting fluctuations for $q$ less then 
 the inverse of the superconducting coherence length.
The magnetic field behavior of
 $\chi(q,0)$  indicates the coexistence of two pseudogaps of
different  origins. One pseudogap dominating $\chi (q, 0)$ near $q=(\pi, \pi)$ is insensitive to  
magnetic fields in our experimental range $\ge 15$ T.  This insensitivity  indicates that
this pseudogap is not intimately tied to superconductivity and that its possible origin
is on a high energy scale which we call a spin-pseudogap.  This nomenclature is motivated 
by the fact that Zeeman contributions to the
excitation spectrum from spin are necessarily much less than $k_B T$ 
in the range of experiments we discuss here. The second  pseudogap, evident in
$\chi (q, 0)$ away from $q \sim (\pi, \pi)$ has a low field scale of $< 10$ T 
and likely originates from superconducting fluctuations as a precursory 
effect of superconductivity.  The latter can be expected since the appropriate 
field scale in this case is determined \cite{Eschrig99} by
 the thermodynamic critical field, $\approx$ 5 T.

Finally, we emphasize that the temperature dependence of all the rates we have 
measured in the  high field limit, changes  
markedly above $T_c$ around $\sim$ 100 - 110 K indicating 
sensitivity to opening of the spin-pseudogap.

We thank  M. Eschrig, J. A. Sauls, D. Rainer for useful discussions, and
 especially  M. Eschrig for help with calculations.
This work is supported by the  National Science Foundation (DMR 91-20000) through the Science and 
Technology Center for Superconductivity. The work at the National High
Magnetic Field Laboratory was supported by the National Science Foundation 
under Cooperative Agreement No. DMR95-27035 and the State of Florida.

%%%%%%%%%%%%%%%%%%%%%%%%%%%%%%%%%%%%%%%%%%%%%%%%%%%%%%%%%%%%%
%%%%%%%%%%%%%%%%%%%%%% APPENDIXES %%%%%%%%%%%%%%%%%%%%%%%%%%%
%%%%%%%%%%%%%%%%%%%%%%%%%%%%%%%%%%%%%%%%%%%%%%%%%%%%%%%%%%%%%

%\vspace{+1cm}
\appendix
\section{Form Factors}
\label{FormFact}

The form factors, relevant for this work, are Shastry-Mila-Rice\cite{Mila89}  
form factors given by,
\begin{eqnarray}
\label{FF_Eq}
& \displaystyle
^{63}F_c = [A_{ab} + 2B(cosq_xa + cosq_ya)]^2
& \nonumber \\
& \displaystyle
^{63}F_{eff} = [A_{c} + 2B(cosq_xa + cosq_ya)]^2
& \nonumber \\
& \displaystyle
^{17}F_ab = 2C^2[cos(q_xa / 2)^2 +cos(q_ya / 2)^2]& \nonumber \\
& \displaystyle
 ^{17-63}F_c = ^{63}F_{eff} * ^{17}F_ab,
& \! \! \! \! \! \! \! \!
\end{eqnarray}
where $A_{ab} = 0.84 B$, $A_{c} = -4 B$, $C = 0.91 B$, and $B = 3.82 * 10^{-7}$ eV.
Their $q$-dependence is shown in \mbox{Fig. \ref{FFact}} along with the imaginary part 
of the susceptibility which is dominated by AF-spin 
fluctuations.

\begin{figure}[h]
%%%%%%%%%%%%%%%%%%%   F I G U R E   %%%%%%%%%%%%%%%%%%%%
\centerline{\epsfxsize0.93\hsize\epsffile{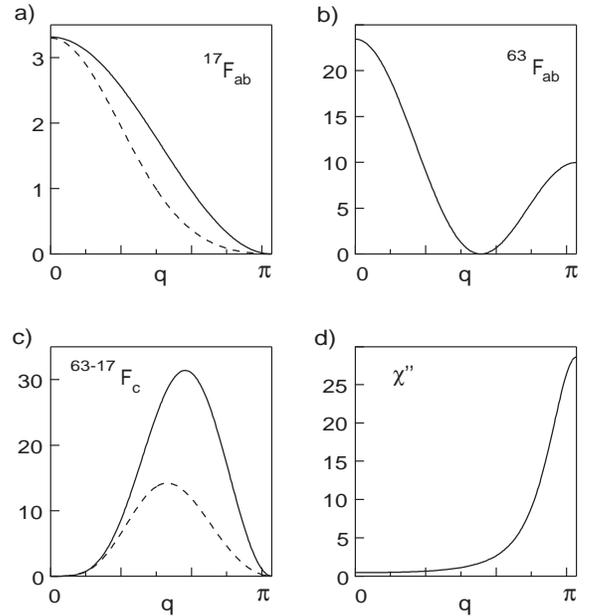}}
%%%%%%%%%%%%%%%%%%%%%%%%%%%%%%%%%%%%%%%%%%%%%%%%%%%%%%%%
\begin{minipage}{0.95\hsize}
\caption[]{\label{FFact}\small 
Form factors\cite{Mila89} in units of $B$, as defined 
in \mbox{Eq. \ref{FF_Eq}} (solid curves) or by \mbox{Eq. \ref{FFOnew_Eq}} (dashed curves):
a) Form factor that determines $^{17}T_1$ for a magnetic field $H_0 || \hat c $. 
b) Form factor that determines $^{63}T_1$ for $H_0 || \hat c $.
c) Form factor that determines $(^{63-17}T_{2})_{ind}$ for $H_0 || \hat c $.
d) $\chi''$ dominated by antiferromagnetic-spin fluctuations 
\cite{millis90} plus a small Fermi-liquid background. 
}
\end{minipage}
\end{figure}
\noindent

The exact form of the oxygen form factor is not generally accepted.
In order to assure a near perfect cancellation of the influence 
of the incommensurate spin-fluctuation peaks (observed by 
neutron scattering \cite{Dap98}) on the $^{17}$O relaxation rates,
 Zha {\it et al.} \cite{Zha96}  suggested that
the oxygen form factor should be altered to include coupling of 
$^{17}$O nuclei to  both nearest-neighbor and next-nearest-neighbor 
Cu$^{2+}$ spins. 
In this case the oxygen form factor, $^{17}F_c$, includes extra terms which they take to be
\begin{eqnarray}
\label{FFOnew_Eq}
& \displaystyle
{ 2C^2 \over (1 + 2r_c)} 
  \{ {cos(q_xa / 2)^2 
[\zeta_\bot
(1 + 2r) -2r +
2rcosq_ya]^2}
& \nonumber \\
& \displaystyle
+  {cos(q_ya / 2)^2 [\zeta_{||}(1 + 2r) -2r + 2rcosq_xa]^2} \},
& \! \! \! \! \! \! \! \!
\end{eqnarray}
\noindent where $r \equiv 0.25$, $\zeta_\bot = 0.91$, and $\zeta_{||} = 1.42$.

%\vspace{-0.5cm}
\bibliographystyle{unsrt}

\vspace{0.5cm}
\end{multicols}
\end{document}